# Performance Evaluation of Two-Hop Wireless Link under Nakagami-m Fading

Afsana Nadia[1], Arifur Rahim Chowdhury[2], Md. Shoayeb Hossain[3] and M. R. Amin[5]

Dept. of Electronics and Communication Engineering
East West University, Dhaka, Bangladesh

Md. Imdadul Islam[4]

Department of Computer Science and Engineering
Jahangirnagar University
Dhaka, Bangladesh

*Abstract*—Now-a-days, intense research is going on two-hop wireless link under different fading conditions with its remedial measures. In this paper work, a two-hop link under three different conditions is considered: (i) MIMO on both hops, (ii) MISO in first hop and SIMO in second hop and finally (iii) SIMO in first hop and MISO in second hop. The three models used here give the flexibility of using STBC (Space Time Block Coding) and combining scheme on any of the source to relay (S-R) and relay to destination (R-D) link. Even incorporation of Transmitting Antenna Selection (TAS) is possible on any link. Here, the variation of SER (Symbol Error Rate) is determined against mean SNR (Signal-to-Noise Ratio) of R-D link for three different modulation schemes: BPSK, 8-PSK and 16-PSK, taking the number of antennas and SNR of S-R link as parameters under Nakagami -m fading condition.

## I. INTRODUCTION

Most common protocols used in wireless networks are the decode-and-forward (DF) and amplify-and-forward (AF) methods [1]. AF protocol used the knowledge of the instantaneous channel state information (CSI) of the source to relay (S-R) channel to control the gain by the relay [2]. The CSI assisted relay may also use the knowledge of the CSI of the S-R channel in the gain; we refer to this as CSI-assisted AF (CSIAF). When the direct link between source and destination is neglected, this is referred to a two-hop network. Two-hop relay networks, where the channel from source to destination (S-D) is split into two possibly shorter links using a relay, are attractive when the direct link between the source and destination is in deep fading. The performance analysis of the two-hop relay network has gained a lot of attention [3]. The two-hop relaying techniques are one of the promisingwireless technologies that have been kindling an enormous interest from the wireless community in the last decade [4]. To achieve higher reliability and throughput for wireless networks, half-duplex two-way relaying system has attracted much research interest [5]. Two-hop relaying communication has a number of advantages over direct-link transmission in terms of connectivity, power saving and channel capacity for the high data-rate coverage required for future cellular and ad-hoc networks [6]. Relaying is a convenient solution to satisfy the requirements of the next generation wireless communication systems, such as: high data rates and large coverage areas [7].

Multiple-antenna systems, also known as multiple-input multiple-output (MIMO) radio, can improve the capacity and reliability of radio communication. In this system, the multiple antenna elements are at both the transmitter and the receiver [8]. They were first investigated by computer simulations in the 1980s [9], and later papers explored them analytically. Since that time, interest in MIMO systems has exploded. The multiple antennas in MIMO systems can be exploited in two different ways. One is the creation of a highly effective antenna diversity system; the other is the use of the multiple antennas for the transmission of several parallel data streams to increase the capacity of the system. The multiple antennas also increase the average SNR seen at the combined output [10]. Antenna diversity is used in wireless systems to combat the effects of fading. If multiple independent copies of the same signal are available, we can combine them to a total signal with high quality, even if some of the copies exhibit low quality.

Deploying multiple antennas in wireless relay networks, referred to as MIMO relaying, has been identified as a promising technique to combat fading and increase transmission reliability [11]–[12]. Transmit antenna selection (TAS) with receive maximal-ratio combining (MRC) in MIMO relaying was proposed in [13]. In TAS/MRC relaying, a single antenna is selected at the transmitters and all the antennas at the receivers are MRC combined [14]. In [15], authors presented a framework for the comparative analysis of TAS/MRC and TAS with receive selection combining (TAS/SC) in a two-hop AF relay network. TAS/MRC and TAS/SC are two attractive MIMO protocols.

This paper presents the performance of a two-hop link where the number of antennas and SNR of S-R as parameters is evaluated under Nakagami-m fading environment separately for BPSK, 8-PSK and 16-PSK modulation schemes. The objective of the paper is to observe the relative impact of Nakagami-m fading environment on the two above mentioned modulation schemes under three different conditions; MIMO on the both hop, MISO in the first hop and SIMO in the second hop, SIMO in the first hop and MISO in the second hop. Though Nakagami-m fading affects the two-hop link, hence some additional techniques like: adaptive equalization, combining scheme of MIMO, incorporation of space-time block code (STBC) etc. are recommended to enhance the performance of such links.

The rest of the paper is organized as follows. Section II describes the system model under consideration. In section III, performance analysis and results from the system model are presented. Finally, section IV concludes the paper.

## II. SYSTEM MODEL

We consider a two-hop wireless network as shown in Fig. 1, where the source node $S$, communicates with the destination node $D$, through the relay node $R$ and they equipped with $N_S$, $N_R$, $N_D$ antennas respectively. There is no direct link between $S$ and $D$ and the communication can be performed only through the relay $R$. This introduces fixed gain on the received signal regardless of the amplitude on the first hop, hence in an output signal with variable power, this type of fixed gain relay is cost effective to implement. Here, space diversity technique is used. Space diversity technique employs multiple transmit or receive antennas having some separation between the adjacent antennas [16]. Various techniques are available to combine the signals from multiple diversity branches. As has been mentioned in the introduction that MRC scheme is one of them. MRC represents a theoretically optimal combiner over fading channels as a diversity scheme in a communication system. Theoretically, multiple copies of the same information signal are combined so as to maximize the instantaneous SNR at the output [17]. Here $N_S= i$, $N_R = j$ and $N_D = k$ where $i = 1,2,…n$; $j = 1,2,…n$; and $k = 1,2,…n$.

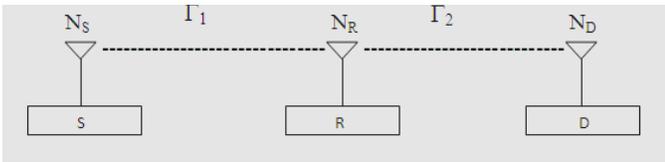

Fig.1.    A Two-Hop Wireless Link.

Let $\Gamma_1$ and $\Gamma_2$ are the random variables representing the SNRs of link S-R and R-D. Then, the equivalent SNR will be

$$\Gamma_{eq} = \frac{\Gamma_1\Gamma_2}{\Gamma_1 + \Gamma_2 + 1} \qquad (1)$$

Let the individual probability density function (PDF) and cumulative distribution function (CDF) of S-R and R-D links are $f_{\Gamma_1}(\gamma_1)$, $F_{\Gamma_1}(\gamma_1)$, $f_{\Gamma_2}(\gamma_2)$, $F_{\Gamma_2}(\gamma_2)$ respectively. Then, the CDF of the equivalent link will be [18]

$$F_{\Gamma_{eq}}(\gamma) = \int_0^\infty Pr\{\gamma_{eq} \le \gamma | \gamma_2\} \, f_{\Gamma_2}(\gamma_2) d\gamma_2.$$

Which gives, $F_\Gamma \;_{eq} \; (\gamma) = \int_0^\infty Pr \; \{ \frac{\gamma_1\gamma_2}{\gamma_1+\gamma_2} \le \gamma | \gamma_2\} f_{\Gamma_2}(\gamma_2) d\gamma_2$ (2)

Now, with the condition, $\frac{\gamma_1\gamma_2}{\gamma_1+\gamma_2+1} \le \gamma$,

We have, $\gamma_1\gamma_2 \le \gamma\gamma_1 + \gamma\gamma_2 + \gamma$

Or, $\gamma_1(\gamma_2 - \gamma) \le \gamma(\gamma_2 + 1)$

Or, $\gamma_1 \le \frac{\gamma(\gamma_2+1)}{\gamma_2-\gamma}$ (3)

For the condition, when $\gamma_2 > \gamma$, we have

$\gamma_1 \le \frac{\gamma(\gamma_2+1)}{\gamma_2-\gamma}$; the range of $\gamma_2$ will be $[\gamma, \infty]$ (4)

Again, when $\gamma_2 < \gamma$,

$\gamma_1 \le \frac{\gamma(\gamma_2+1)}{\gamma_2-\gamma}$; the range of $\gamma_2$ will be $[o, \gamma]$ (5)

From (1),

$$F_{\Gamma_{eq}}(\gamma) = \int_0^\gamma Pr\{\gamma_1 \ge \frac{\gamma(\gamma_2+1)}{\gamma_2-\gamma} | \gamma_2\} \, f_{\Gamma_2}(\gamma_2) d\gamma_2$$

$$+ \int_\gamma^\infty Pr\{\gamma_1 \le \frac{\gamma(\gamma_2+1)}{\gamma_2-\gamma} | \gamma_2\} f_{\Gamma_2}(\gamma_2) d\gamma_2$$

Or, $F_{\Gamma_{eq}}(\gamma) = I_1 + I_2$ (6)

Where $I_1 = \int_0^\gamma Pr\{\gamma_1 \ge \frac{\gamma(\gamma_2+1)}{\gamma_2-\gamma} | \gamma_2\} \, f_{\Gamma_2}(\gamma_2) d\gamma_2$

and

$I_2 = \int_\gamma^\infty Pr\{\gamma_1 \le \frac{\gamma(\gamma_2+1)}{\gamma_2-\gamma} | \gamma_2\} \, f_{\Gamma_2}(\gamma_2) d\gamma_2$

We know, the cdf will be

$F_x(x) = \int_0^x f_x(x) \, dx$ (7)

and, $Pr\{x \le r\} = F_x(r)$ (8)

Then, $I_1 = \int_0^\gamma Pr\{\gamma_1 \ge \frac{\gamma(\gamma_2+1)}{\gamma_2-\gamma} | \gamma_2\} \, f_{\Gamma_2}(\gamma_2) d\gamma_2$

$= \int_0^\gamma \{1 - F_{\Gamma_1}(\frac{\gamma(\gamma_2+1)}{\gamma_2-\gamma})\} f_{\Gamma_2}(\gamma_2) d\gamma_2$

$= \int_0^\gamma [1 - \{1 - e^{-\frac{\gamma(\gamma_2+1)}{\gamma_2-\gamma} \cdot \frac{1}{\overline{\gamma_1}}}\}] f_{\Gamma_2}(\gamma_2) d\gamma_2$

$= \int_0^\gamma e^{-\frac{\gamma(\gamma_2+1)}{\gamma_2-\gamma} \cdot \frac{1}{\overline{\gamma_1}}} \frac{1}{\overline{\gamma_2}} e^{-\frac{\gamma_2}{\overline{\gamma_2}}} d\gamma_2$ (9)

And $I_2 = \int_\gamma^\infty Pr\{\gamma_1 \le \frac{\gamma(\gamma_2+1)}{\gamma_2-\gamma} | \gamma_2\} \, f_{\Gamma_2}(\gamma_2) d\gamma_2$

$= \int_\gamma^\infty F_{\Gamma_1}\{\frac{\gamma(\gamma_2+1)}{\gamma_2-\gamma}\} f_{\Gamma_2}(\gamma_2) d\gamma_2$

$= \int_\gamma^\infty \{1 - e^{-\frac{\gamma(\gamma_2+1)}{\gamma_2-\gamma} \cdot \frac{1}{\overline{\gamma_1}}}\} \frac{1}{\overline{\gamma_2}} e^{-\frac{\gamma_2}{\overline{\gamma_2}}} d\gamma_2$ (10)

Using (6), (9) and (10), we can write

$F_{\Gamma_{eq}}(\gamma) = \int_0^\gamma e^{-\frac{\gamma(\gamma_2+1)}{\gamma_2-\gamma} \cdot \frac{1}{\overline{\gamma_1}}} \frac{1}{\overline{\gamma_2}} e^{-\frac{\gamma_2}{\overline{\gamma_2}}} d\gamma_2$

$+ \int_\gamma^\infty \{1 - e^{-\frac{\gamma(\gamma_2+1)}{\gamma_2-\gamma} \cdot \frac{1}{\overline{\gamma_1}}}\} \frac{1}{\overline{\gamma_2}} e^{-\frac{\gamma_2}{\overline{\gamma_2}}} d\gamma_2$ (11)

The symbol error rate (SER) [15] will be

$P_{SER} = \frac{a}{2} \cdot \sqrt{\frac{b}{\pi}} \int_0^\gamma \frac{e^{-b\gamma}}{\sqrt{\gamma}} F_{\Gamma_{eq}}(\gamma) \, d\gamma$ (12)

Where, *a* and *b* are the constellation-specific constants.

In the following, we have considered three cases depending on the number of antenna of source, relay and destination.

### A. Case A: Mimo on Both Hops

In the case A, we consider a distributed wireless network where node *S* is equipped with a multiple transmitting antennas $S_1$, $S_2$ and $S_3$ and node *D* is equipped with a multiple receiving antennas $D_1$, $D_2$ and $D_3$ whereas the relay has three antennas $R_1$, $R_2$ and $R_3$ in Fig. 2. These three antennas $R_1$, $R_2$ and $R_3$ play the role of receiving antennas when they receive signals from the transmitting antennas $S_1$, $S_2$ and $S_3$. These three antennas $R_1$, $R_2$ and $R_3$ also play the role of transmitting antennas when they transmit signals to the receiving antennas $D_1$, $D_2$ and $D_3$.

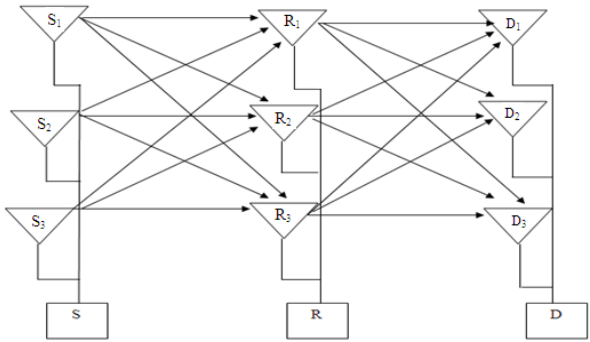

Fig.2.        MIMO on Both Hops

### B. Case B: MISO in the First Hop and SIMO in the Second Hop

In the case B, node *S* is equipped with a multiple transmitting antennas $S_1$, $S_2$ and $S_3$ and node *D* is equipped with a multiple receiving antennas $D_1$, $D_2$ and $D_3$ whereas the relay has a single antenna $R_1$ in Fig. 3. The antenna $R_1$ plays the role of receiving antenna when it receives signals from the transmitting antennas $S_1$, $S_2$ and $S_3$. The antenna $R_1$ also plays the role of transmitting antenna when it transmits signals to the receiving antennas $D_1$, $D_2$ and $D_3$. Therefore, from the path S to R, space diversity is used as there are three transmitters and one receiver. Now, from R to D, the method of MRC is used as there are only one transmitter and three receivers.

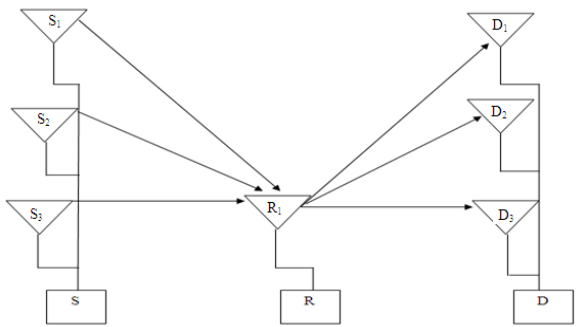

Fig.3.        MISO in the First Hop and SIMO in the Second Hop

### C. Case C: SIMO in the First Hop and MISO in the Second Hop

In the case C, node *S* is equipped with a single transmitting antenna $S_1$ and node *D* is equipped with a single receiving antenna $D_1$ whereas the relay has three antennas $R_1$, $R_2$ and $R_3$ in Fig. 4. These three antennas $R_1$, $R_2$ and $R_3$ play the role of receiving antennas when they receive signals from the transmitting antenna $S_1$. These three antennas $R_1$, $R_2$ and $R_3$ also play the role of transmitting antennas when they transmit signals to the receiving antenna $D_1$. Therefore, from the path S to R, the method MRC is used as there are only one transmitter and three receivers; from R to D, orthogonal scheme is used as there are three transmitters and one receiver.

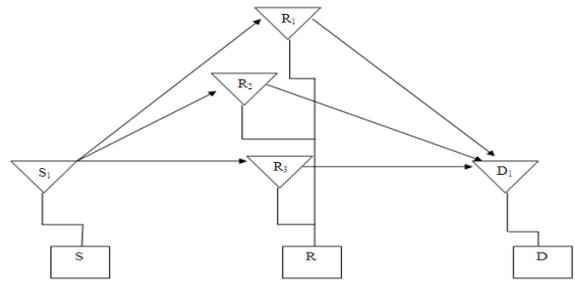

Fig.4.        SIMO in the First Hop and MISO in the Second Hop

### III. RESULTS

Fig. 5 shows the variation of SER against the instantaneous SNR of R-D link taking $N_S = N_R = N_D = 3$. The following modulation schemes: BPSK, 8- PSK and 16- PSK are considered taking SNR between S-R link as 3dB and 2 dB. Here, we see from Fig. 5 that the SER decreases very rapidly till 5 dB of the mean SNR between R-D. After 5 dB improvement is very slow. For any further improvement, we have to increase the SNR between the source and the relay as visualized form Fig. 5.

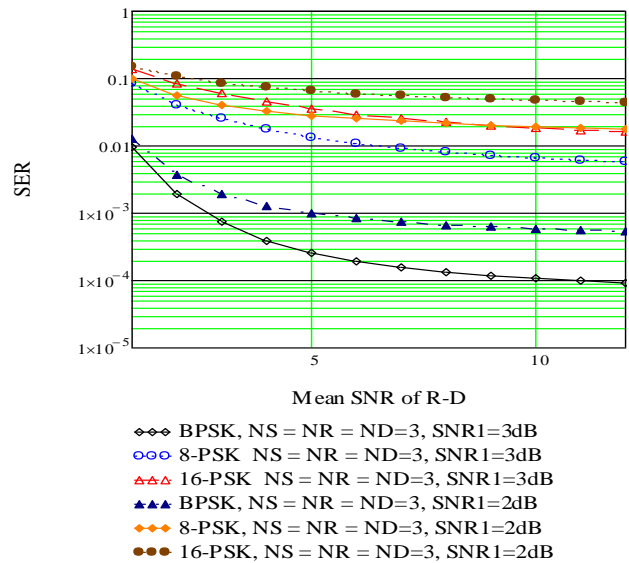

◇◇◇  BPSK, NS = NR = ND=3, SNR1=3dB
○○○  8-PSK  NS = NR = ND=3, SNR1=3dB
△△△  16-PSK  NS = NR = ND=3, SNR1=3dB
▲▲▲  BPSK, NS = NR = ND=3, SNR1=2dB
✦✦✦  8-PSK, NS = NR = ND=3, SNR1=2dB
●●●  16-PSK, NS = NR = ND=3, SNR1=2dB

Fig.5.        Impact of Modulation Scheme and SNR of link-1

The impact of number of antennas is visualized from Fig. 6. Similar to the previous case like Fig. 5, it is observed that the SER decreases with the mean SNR. Here, the parameter of the curves is the number of antennas of MIMO link. Incorporation of a signal antenna at each step increases the performance of the SER tremendously. The impact of the number of antennas is more prominent of BPSK scheme than the other two modulation schemes.

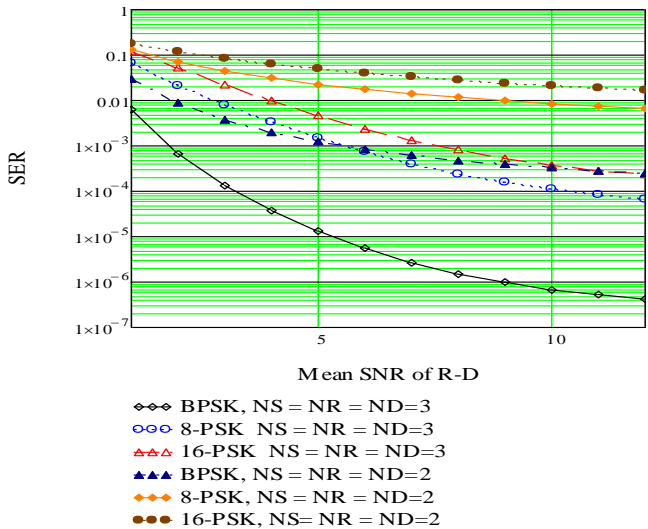

Fig.6.     Impact of Modulation Scheme and the number of antennas

Next, we use MISO technique in the first hop and SIMO technique in the second hop so that the space diversity can be applied at the first hop and MRC can be used at the second hop. In this case, performance is heavily improved with the incorporation of one additional antenna at the source and the destination as shown in Fig. 7. But the overall performance of such a scheme is inferior to the MIMO case.

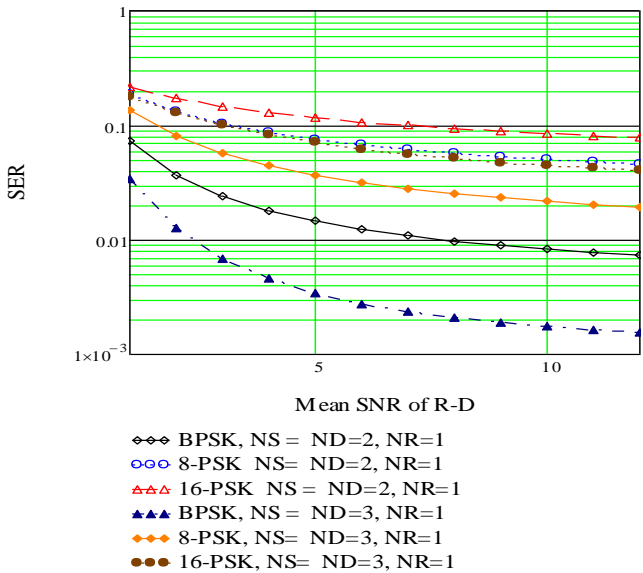

Fig.7.     Impact of Modulation Scheme, numbers of transmitting and receiving antennas

Finally, we use single antenna at both the source and the destination but multiple antennas on the relay, hence MRC and orthogonal scheme can also be applied in this case. Fig. 8 shows the performance of the scheme for $N_R$= 2 and $N_R$= 4 cases. This scheme gives marginally better performance than that of the case of Fig. 7 for the same number of antennas and modulation schemes.

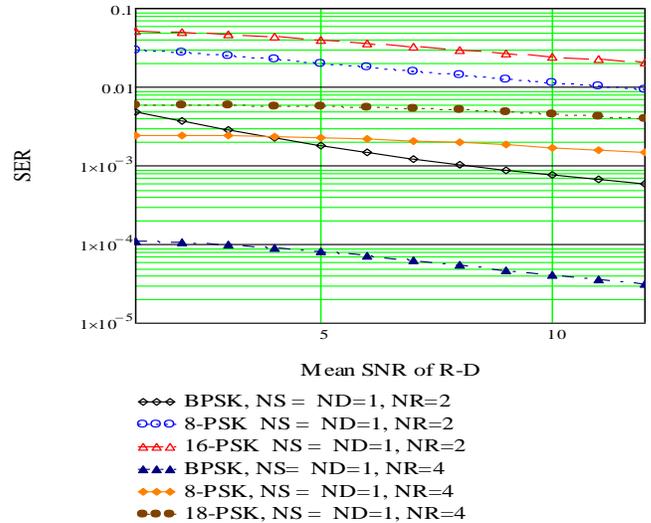

Fig.8.     Impact of Modulation Scheme and numbers of relaying antennas

## IV.   CONCLUSION

The finding of the paper is that, instead of using multiple antennas at the sender and the receiver one can use multiple antennas only on the relay but single antenna at the sender and the receiver. Therefore, SIMO in the first hop and MISO in the second hop is better scheme than MISO in the first hop and SIMO in the second hop case. Although, MIMO on the both hops is better than anyone of above model. But if we use SIMO-MISO scheme, cost could be minimized. So, SIMO-MISO combination is the best in context of SER and cost. We have used this technique only for Nakagami-m fading. In future, the above phenomenon can also be observed for Rayleigh, Rician and K-fading cases. Furthermore, equalizer could be used to improve the overall system performance. This work is under investigation and will be reported soon in future.

AUTHOR PROFILE

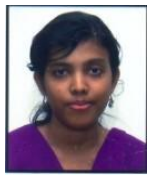

**Afsana Nadia** has completed her B.Sc. Engineering in Electronics and Telecommunication Engineering from Daffodil International University, Dhaka, Bangladesh in 2011 and M.S. Engineering in Telecommunication Engineering from East West University, Dhaka, Bangladesh in 2012. She is the associate member of Bangladesh Computer Society.

Her fields of research are computer networks, wireless and mobile communications.

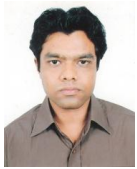

**Arifur Rahim Chowdhury** has completed his B.Sc. in Electronics and Telecommunication Engineering degree from the People's University of Bangladesh, Dhaka in 2011 and M.S. Engineering in Telecommunication Engineering from East West University, Dhaka, Bangladesh in 2012. His field of study is wireless communications and networks.

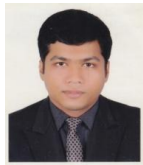

**Md. Shoayeb Hossain** has completed his B.Sc. in Information and Telecommunication Engineering degree from Darul Ihsan University, Dhaka, Bangladesh in 2010 and M.S. Engineering in Telecommunication Engineering from East West University, Dhaka, Bangladesh in 2012. His field of study is wireless and mobile communications.

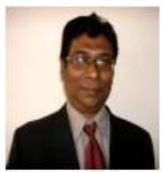

**Md. Imdadul Islam** has completed his B.Sc. and M.Sc Engineering in Electrical and Electronic Engineering from Bangladesh University of Engineering and Technology, Dhaka, Bangladesh in 1993 and 1998 respectively and has completed his Ph.D degree from the Department of Computer Science and Engineering, Jahangirnagar University, Dhaka, Bangladesh in the field of network traffic engineering in 2010.

He is now working as a Professor at the Department of Computer Science and Engineering, Jahangirnagar University, Savar, Dhaka, Bangladesh. Previously, he worked as an Assistant Engineer in Sheba Telecom (Pvt.) LTD (A joint venture company between Bangladesh and Malaysia, for Mobile cellular and WLL), from Sept.1994 to July 1996. Dr Islam has a very good field experience in installation of Radio Base Stations and Switching Centers for WLL. His research field is network traffic, wireless communications, wavelet transform, OFDMA, WCDMA, adaptive filter theory, ANFIS and array antenna systems. He has more than hundred research papers in national and international journals and conference proceedings.

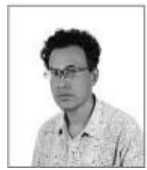

**M. R. Amin** received his B.S. and M.S. degrees in Physics from Jahangirnagar University, Dhaka, Bangladesh in 1984 and 1986 respectively and his Ph.D. degree in Plasma Physics from the University of St. Andrews, U. K. in 1990. He is currently working as a Professor of Electronics and Communications Engineering at East West University, Dhaka, Bangladesh. He served as a Post-Doctoral Research Associate in Electrical Engineering at the University of Alberta, Canada, during 1991-1993. He was an Alexander von Humboldt Research Fellow at the Max-Planck Institute for Extraterrestrial Physics at Garching/Munich, Germany during 1997-1999.

Dr. Amin was awarded the Commonwealth Postdoctoral Fellowship in 1997. Besides these, he has also received several awards for his research, including the Bangladesh Academy of Science Young Scientist Award for the year 1996 and the University Grants Commission Young Scientist Award for 1996. His current research fields are wireless communications and networks and also nonlinear plasma dynamics. He is a member of the IEEE.